\documentclass[sigconf]{acmart}

\usepackage{tikz}
\usepackage{pgfplots}
\pgfplotsset{compat=1.18}
\usepackage{array}
\usepackage[caption=false,font=normalsize,labelfont=sf,textfont=sf]{subfig}
\usepackage{stfloats}
\usepackage{url}
\usepackage{verbatim}
\usepackage{graphicx}
\usepackage{multirow}
\usepackage{textcomp}
\usepackage{algorithm}
\usepackage{algorithmic}
\usepackage{cases}
\usepackage{bm}
\usepackage{color,xcolor}
\usepackage{booktabs} 
\usepackage{subfig}
\usepackage{balance}
\usepackage{setspace}
\usepackage{threeparttable}
\usetikzlibrary{patterns}
\allowdisplaybreaks[2]

\AtBeginDocument{%
  }

\copyrightyear{2026}
\acmYear{2026}
\setcopyright{cc}
\setcctype{by}
\acmConference[ICCAD '26]{IEEE/ACM International Conference on Computer-Aided Design}{November 08--12, 2026}{San Jose, CA, USA}
\acmBooktitle{IEEE/ACM International Conference on Computer-Aided Design (ICCAD '26), November 08--12, 2026, San Jose, CA, USA}
\acmDOI{10.1145/3831252.3834205}
\acmISBN{979-8-4007-2873-0/2026/11}

\begin{document}
\title[Near-Optimal Floorplanning with Complex Design Constraints]{
Toward Near-Optimal Floorplanning with Complex Design Constraints for Modern SoCs
}

\author{\texorpdfstring{
Fuxing Huang\textsuperscript{1}, 
Hao Ai\textsuperscript{2},
Qiyuan Chen\textsuperscript{1}, 
Hao Wu\textsuperscript{1}, 
Wenxing Zhu\textsuperscript{2},
Xinning Liu\textsuperscript{1},
Ziran Zhu\textsuperscript{1,*}
}{Fuxing Huang, Hao Ai, Qiyuan Chen, Hao Wu, Wenxing Zhu, Xinning Liu, Ziran Zhu}}
\affiliation{
\textsuperscript{1}School of Integrated Circuits, Southeast University, Nanjing, China\\
\textsuperscript{2}School of Mathematics and Statistics, Fuzhou University, Fuzhou, China\\
\country{}
}
\thanks{This work was partially supported by the National Natural Science Foundation of China under Grant 92373207, and the Zhishan Young Scholar Support Program of Southeast University under Grant 2242024RCB0052. (*Corresponding author: zrzhu@seu.edu.cn)}

\renewcommand{\shortauthors}{Fuxing Huang et al.}

\begin{abstract}Modern system-on-chip (SoC) floorplanning is increasingly challenged by highly coupled  complex design constraints. Existing methods often relax constraints into objectives via multi-objective optimization, leading to infeasible solutions or degraded quality. To address this, we formulate the problem as a constrained optimization task and propose a novel three-stage analytical framework integrating semidefinite programming (SDP), augmented Lagrangian method (ALM), and second-order cone programming (SOCP), which enforces constraints more rigorously and achieves higher-quality solutions.
First, we adopt a two-step SDP formulation to obtain a robust initialization in the non-convex search space. Next, we incorporate differentiable objectives and constraints into an incremental ALM to jointly handle coupled design constraints during global floorplanning. Finally, an SOCP-based legalization strictly enforces feasibility while further reducing area.
Experiments on the FloorSet benchmark show that our method satisfies all constraints with the fastest runtime. Compared with state-of-the-art methods, it reduces wirelength and area by 90\% and 16\% on average, and achieves near-optimal results, with wirelength and area being merely 2\% and 1\% larger than the ground-truth references, respectively.
\end{abstract}

\maketitle

\section{Introduction}
The continuous advancement of semiconductor technology has enabled modern SoCs to integrate billions of transistors, drastically escalating overall design complexity. To manage this unprecedented scale and optimize power, performance, and area (PPA), hierarchical design methodologies and intellectual property (IP) reuse have become indispensable. Within this modular paradigm, top-level floorplanning serves as the foundation that directly dictates successful system integration and ultimate chip performance~\cite{tseng2024challenges,hosny2021characterizing,StudyFloorplanning}.

\begin{figure}
  \includegraphics[width=0.95\hsize]{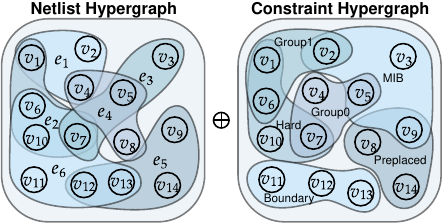}
  \caption{The challenge of modern constrained floorplanning.}
  \label{fig:intro}
  \vspace{-3mm}
\end{figure}

Modern constrained floorplanning builds upon traditional floorplanning by incorporating more complex coupled design constraints, as shown in Figure ~\ref{fig:intro}. These constraints are fundamental requirements for ensuring design validity and engineering feasibility, rather than merely minor adjustments. Recent industrial benchmarks and contest specifications also emphasize that modern evaluations should treat these constraints in an integrated manner, rather than treating them in isolation, while optimizing wirelength and area~\cite{2025FloorSetChallenge,floorset}. In this context, the key challenge is how to effectively and efficiently handle these complex design constraints while optimizing the wirelength and area objectives. 

Previous research generally falls into three main categories: metaheuristic, learning-based, and analytical methods. 
Metaheuristic methods~\cite{1986Stree,1995SP,2000BTree,2000CBL,young2004cdfp,2024PARSAC,li2025ftafp,2003Parquet,2012GA}, which rely on iterative search and local perturbations, have been widely used for floorplanning due to their flexibility in handling various constraints. However, they typically rely on packing-based representations to explicitly enforce non-overlap constraints, while incorporating other constraints into the cost function as soft penalties. Even recent state-of-the-art approaches such as PARSAC~\cite{2024PARSAC} utilize parallel search to handle these constraints. However, they still fundamentally struggle to guarantee strict satisfaction as the constraint scale grows.

Learning-based methods~\cite{2020rlsa,nature,liu2022MLFPDAC,2022goodfp,2022kdd,zhong2024flexplanner}, particularly reinforcement learning, have demonstrated exceptional potential for rapid inference on specific data distributions by using rewards to guide agents.
However, training these models requires significant computational time, and their generalization capabilities on unseen designs remain limited~\cite{cheng2023assessment,cheng2025updatedassessment}. Furthermore, introducing complex design constraints during the learning process drastically increases the difficulty of training convergence, which inevitably further weakens their generalization capabilities.

Analytical methods~\cite{2006Analytical,2008AR,2011UFO,2014F-FM,2023SDP,2023rotationfp,2024NTUFP,remap,2024jigsawplanner,sdp3dfp} formulate floorplanning as continuous optimization and currently achieve leading performance in both quality and efficiency. Nevertheless, these approaches are primarily designed for basic overlap removal and wirelength optimization within a standard two-stage framework. They lack a systematic mechanism to accommodate highly coupled design constraints. While one could append these complex constraints as soft penalty terms to existing global analytical engines, this would inevitably trap the solver in poor local optima. Moreover, the subsequent local legalization stage would struggle to achieve strictly satisfies the constraints floorplan without catastrophic degradation in solution quality.

In summary, existing methods treat complex design constraints merely as penalties, which frequently fails to ensure satisfies the constraints. Therefore, to effectively handle these complex constraints, we propose a rigorous analytical framework. Our main contributions are summarized as follows:
\begin{itemize}
  \item To the best of our knowledge, this is the first analytical floorplanning method that integrates SDP, ALM, and SOCP into a unified, end-to-end framework capable of handling complex design constraints.
  \item To reliably initialize the floorplan within a highly non-convex search space, we propose an efficient two-step SDP initialization. This approach robustly establishes a high-quality initial solution, effectively avoiding poor local minima.
  \item We develop differentiable models for both objectives and constraints, enabling their direct integration into an incremental ALM. This strategy seamlessly transitions from the initial solution to continuous optimization, simultaneously resolving multiple coupled constraints while ensuring stable convergence and high-quality solutions.
  \item To guarantee strict final legality, we construct a constraint graph capturing the global floorplanning results and formulate the legalization as an SOCP. This formulation rigorously eliminates all remaining constraint violations while further minimizing the floorplan area.
  \item The results of the FloorSet benchmark demonstrate that our method achieves the fastest runtime while satisfying all constraints. Compared to state-of-the-art methods, it reduces HPWL by 90\% and area by 16\% on average. Furthermore, it achieves near-optimal results, with wirelength and area being merely 2\% and 1\% larger than the ground-truth references, respectively.
\end{itemize}

\section{Constrained Floorplanning Problem}\label{sec:pre}
Given a set of $n$ blocks $\mathbf{V}=\{v_1,\dots,v_n\}$ and a set of $m$ nets $\mathbf{E}=\{e_1,\dots,e_m\}$, constrained floorplanning aims to determine the shape and placement of all blocks.
Each block $v_i$ is represented by center coordinates $(x_i,y_i)$ and dimensions $(w_i,h_i)$.
The optimization variables are collectively denoted by $\boldsymbol{v}=\{x_i,y_i,w_i,h_i \mid i=1,\dots,n\}$.

\textbf{Objectives.} 
The objective combines two quality metrics: wirelength and floorplan area.
Wirelength is estimated by the weighted half-perimeter wirelength (HPWL):
\begin{equation*}
    \mathrm{HPWL}(\boldsymbol{v})=
    \sum_{e_i\in\mathbf{E}} \omega_{i} \left(\max_{v_j\in e_i} x_j - \min_{v_j\in e_i} x_j + \max_{v_j\in e_i} y_j - \min_{v_j\in e_i} y_j\right),
\end{equation*}
where $\omega_i$ denotes the weight of net $e_i$.
The floorplan area is defined as the area of the enclosing bounding box:
\begin{equation*}
    \text{Area}(\boldsymbol{v}) = \left( \max_{v_i \in \mathbf{V}} x_i^+ - \min_{v_i \in \mathbf{V}} x_i^- \right) \times \left( \max_{v_i \in \mathbf{V}} y_i^+ - \min_{v_i \in \mathbf{V}} y_i^- \right),
\end{equation*}
where $x_i^\pm = x_i \pm \frac{w_i}{2}$ and $y_i^\pm = y_i \pm \frac{h_i}{2}$ are the block edges.

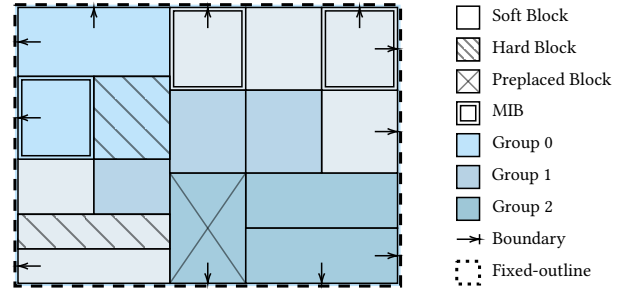
\begin{figure}
    \raggedleft

\tikzset{
pattern size/.store in=\mcSize, 
pattern size = 5pt,
pattern thickness/.store in=\mcThickness, 
pattern thickness = 0.3pt,
pattern radius/.store in=\mcRadius, 
pattern radius = 1pt}
\makeatletter
\pgfutil@ifundefined{pgf@pattern@name@_ydzad1jjf}{
\pgfdeclarepatternformonly[\mcThickness,\mcSize]{_ydzad1jjf}
{\pgfqpoint{0pt}{-\mcThickness}}
{\pgfpoint{\mcSize}{\mcSize}}
{\pgfpoint{\mcSize}{\mcSize}}
{
\pgfsetcolor{\tikz@pattern@color}
\pgfsetlinewidth{\mcThickness}
\pgfpathmoveto{\pgfqpoint{0pt}{\mcSize}}
\pgfpathlineto{\pgfpoint{\mcSize+\mcThickness}{-\mcThickness}}
\pgfusepath{stroke}
}}
\makeatother

\tikzset{
pattern size/.store in=\mcSize, 
pattern size = 5pt,
pattern thickness/.store in=\mcThickness, 
pattern thickness = 0.3pt,
pattern radius/.store in=\mcRadius, 
pattern radius = 1pt}
\makeatletter
\pgfutil@ifundefined{pgf@pattern@name@_e442rm2w0}{
\pgfdeclarepatternformonly[\mcThickness,\mcSize]{_e442rm2w0}
{\pgfqpoint{0pt}{-\mcThickness}}
{\pgfpoint{\mcSize}{\mcSize}}
{\pgfpoint{\mcSize}{\mcSize}}
{
\pgfsetcolor{\tikz@pattern@color}
\pgfsetlinewidth{\mcThickness}
\pgfpathmoveto{\pgfqpoint{0pt}{\mcSize}}
\pgfpathlineto{\pgfpoint{\mcSize+\mcThickness}{-\mcThickness}}
\pgfusepath{stroke}
}}
\makeatother

\tikzset{
pattern size/.store in=\mcSize, 
pattern size = 5pt,
pattern thickness/.store in=\mcThickness, 
pattern thickness = 0.3pt,
pattern radius/.store in=\mcRadius, 
pattern radius = 1pt}
\makeatletter
\pgfutil@ifundefined{pgf@pattern@name@_jhy93hey3}{
\pgfdeclarepatternformonly[\mcThickness,\mcSize]{_jhy93hey3}
{\pgfqpoint{0pt}{-\mcThickness}}
{\pgfpoint{\mcSize}{\mcSize}}
{\pgfpoint{\mcSize}{\mcSize}}
{
\pgfsetcolor{\tikz@pattern@color}
\pgfsetlinewidth{\mcThickness}
\pgfpathmoveto{\pgfqpoint{0pt}{\mcSize}}
\pgfpathlineto{\pgfpoint{\mcSize+\mcThickness}{-\mcThickness}}
\pgfusepath{stroke}
}}
\makeatother
\tikzset{every picture/.style={line width=0.75pt}} %
\resizebox{0.95\hsize}{!}{
\begin{tikzpicture}[x=0.75pt,y=0.75pt,yscale=-1,xscale=1]

\draw  [fill={rgb, 255:red, 191; green, 228; blue, 251 }  ,fill opacity=1 ][dash pattern={on 5.63pt off 4.5pt}][line width=1.5]  (1,1) -- (255,1) -- (255,186.33) -- (1,186.33) -- cycle ;
\draw  [fill={rgb, 255:red, 191; green, 228; blue, 251 }  ,fill opacity=1 ] (3,2.82) -- (103,2.82) -- (103,48.24) -- (3,48.24) -- cycle ;
\draw  [fill={rgb, 255:red, 227; green, 236; blue, 242 }  ,fill opacity=1 ] (3,102.75) -- (53,102.75) -- (53,139.09) -- (3,139.09) -- cycle ;
\draw  [fill={rgb, 255:red, 158; green, 199; blue, 215 }  ,fill opacity=0.8 ] (153,111.84) -- (253,111.84) -- (253,148.18) -- (153,148.18) -- cycle ;
\draw  [fill={rgb, 255:red, 227; green, 236; blue, 242 }  ,fill opacity=1 ] (103,2.82) -- (153,2.82) -- (153,57.33) -- (103,57.33) -- cycle ;
\draw  [fill={rgb, 255:red, 227; green, 236; blue, 242 }  ,fill opacity=1 ] (203,57.33) -- (253,57.33) -- (253,111.84) -- (203,111.84) -- cycle ;
\draw  [fill={rgb, 255:red, 191; green, 228; blue, 251 }  ,fill opacity=1 ] (3,48.24) -- (53,48.24) -- (53,102.75) -- (3,102.75) -- cycle ;
\draw  [fill={rgb, 255:red, 158; green, 199; blue, 215 }  ,fill opacity=0.8 ] (103,111.84) -- (153,111.84) -- (153,184.52) -- (103,184.52) -- cycle ;
\draw  [fill={rgb, 255:red, 183; green, 210; blue, 228 }  ,fill opacity=0.8 ] (53,102.75) -- (103,102.75) -- (103,139.09) -- (53,139.09) -- cycle ;
\draw  [fill={rgb, 255:red, 183; green, 210; blue, 228 }  ,fill opacity=1 ] (3,139.09) -- (103,139.09) -- (103,161.8) -- (3,161.8) -- cycle ;
\draw  [fill={rgb, 255:red, 227; green, 236; blue, 242 }  ,fill opacity=1 ] (103,161.8) -- (3,161.8) -- (3,184.52) -- (103,184.52) -- cycle ;
\draw  [fill={rgb, 255:red, 183; green, 210; blue, 228 }  ,fill opacity=0.8 ] (103,57.33) -- (153,57.33) -- (153,111.84) -- (103,111.84) -- cycle ;
\draw  [fill={rgb, 255:red, 227; green, 236; blue, 242 }  ,fill opacity=1 ] (153,2.82) -- (203,2.82) -- (203,57.33) -- (153,57.33) -- cycle ;
\draw  [fill={rgb, 255:red, 227; green, 236; blue, 242 }  ,fill opacity=1 ] (203,2.82) -- (253,2.82) -- (253,57.33) -- (203,57.33) -- cycle ;
\draw  [fill={rgb, 255:red, 183; green, 210; blue, 228 }  ,fill opacity=0.8 ] (153,57.33) -- (203,57.33) -- (203,111.84) -- (153,111.84) -- cycle ;
\draw  [fill={rgb, 255:red, 158; green, 199; blue, 215 }  ,fill opacity=0.8 ] (153,148.18) -- (253,148.18) -- (253,184.52) -- (153,184.52) -- cycle ;
\draw   (205.5,5.09) -- (250.5,5.09) -- (250.5,55.06) -- (205.5,55.06) -- cycle ;
\draw   (105.5,5.09) -- (150.5,5.09) -- (150.5,55.06) -- (105.5,55.06) -- cycle ;
\draw   (5.5,50.51) -- (50.5,50.51) -- (50.5,100.48) -- (5.5,100.48) -- cycle ;
\draw [color={rgb, 255:red, 0; green, 0; blue, 0 }  ,draw opacity=0.5 ]   (103,111.84) -- (153,184.52) ;
\draw [color={rgb, 255:red, 0; green, 0; blue, 0 }  ,draw opacity=0.5 ]   (153,111.84) -- (103,184.52) ;
\draw  [fill={rgb, 255:red, 191; green, 228; blue, 251 }  ,fill opacity=1 ] (53,48.24) -- (103,48.24) -- (103,102.75) -- (53,102.75) -- cycle ;
\draw  [color={rgb, 255:red, 0; green, 0; blue, 0 }  ,draw opacity=0.5 ][pattern=_ydzad1jjf,pattern size=11.25pt,pattern thickness=0.75pt,pattern radius=0pt, pattern color={rgb, 255:red, 0; green, 0; blue, 0}][line width=0.75]  (53,48.24) -- (103,48.24) -- (103,102.75) -- (53,102.75) -- cycle ;
\draw  [color={rgb, 255:red, 0; green, 0; blue, 0 }  ,draw opacity=0.5 ][fill={rgb, 255:red, 227; green, 236; blue, 242 }  ,fill opacity=1 ][line width=0.75]  (3,139.09) -- (103,139.09) -- (103,161.8) -- (3,161.8) -- cycle ;
\draw    (238,166.35) -- (251,166.35) ;
\draw [shift={(253,166.35)}, rotate = 180] [color={rgb, 255:red, 0; green, 0; blue, 0 }  ][line width=0.75]    (6.56,-1.97) .. controls (4.17,-0.84) and (1.99,-0.18) .. (0,0) .. controls (1.99,0.18) and (4.17,0.84) .. (6.56,1.97)   ;
\draw    (203,170.89) -- (203,182.52) ;
\draw [shift={(203,184.52)}, rotate = 270] [color={rgb, 255:red, 0; green, 0; blue, 0 }  ][line width=0.75]    (6.56,-1.97) .. controls (4.17,-0.84) and (1.99,-0.18) .. (0,0) .. controls (1.99,0.18) and (4.17,0.84) .. (6.56,1.97)   ;
\draw    (238,84.58) -- (251,84.58) ;
\draw [shift={(253,84.58)}, rotate = 180] [color={rgb, 255:red, 0; green, 0; blue, 0 }  ][line width=0.75]    (6.56,-1.97) .. controls (4.17,-0.84) and (1.99,-0.18) .. (0,0) .. controls (1.99,0.18) and (4.17,0.84) .. (6.56,1.97)   ;
\draw    (238,30.07) -- (251,30.07) ;
\draw [shift={(253,30.07)}, rotate = 180] [color={rgb, 255:red, 0; green, 0; blue, 0 }  ][line width=0.75]    (6.56,-1.97) .. controls (4.17,-0.84) and (1.99,-0.18) .. (0,0) .. controls (1.99,0.18) and (4.17,0.84) .. (6.56,1.97)   ;
\draw    (128,170.89) -- (128,182.52) ;
\draw [shift={(128,184.52)}, rotate = 270] [color={rgb, 255:red, 0; green, 0; blue, 0 }  ][line width=0.75]    (6.56,-1.97) .. controls (4.17,-0.84) and (1.99,-0.18) .. (0,0) .. controls (1.99,0.18) and (4.17,0.84) .. (6.56,1.97)   ;
\draw    (53,4.82) -- (53,16.44) ;
\draw [shift={(53,2.82)}, rotate = 90] [color={rgb, 255:red, 0; green, 0; blue, 0 }  ][line width=0.75]    (6.56,-1.97) .. controls (4.17,-0.84) and (1.99,-0.18) .. (0,0) .. controls (1.99,0.18) and (4.17,0.84) .. (6.56,1.97)   ;
\draw    (5,25.53) -- (18,25.53) ;
\draw [shift={(3,25.53)}, rotate = 0] [color={rgb, 255:red, 0; green, 0; blue, 0 }  ][line width=0.75]    (6.56,-1.97) .. controls (4.17,-0.84) and (1.99,-0.18) .. (0,0) .. controls (1.99,0.18) and (4.17,0.84) .. (6.56,1.97)   ;
\draw    (128,4.82) -- (128,16.44) ;
\draw [shift={(128,2.82)}, rotate = 90] [color={rgb, 255:red, 0; green, 0; blue, 0 }  ][line width=0.75]    (6.56,-1.97) .. controls (4.17,-0.84) and (1.99,-0.18) .. (0,0) .. controls (1.99,0.18) and (4.17,0.84) .. (6.56,1.97)   ;
\draw    (5,173.16) -- (18,173.16) ;
\draw [shift={(3,173.16)}, rotate = 0] [color={rgb, 255:red, 0; green, 0; blue, 0 }  ][line width=0.75]    (6.56,-1.97) .. controls (4.17,-0.84) and (1.99,-0.18) .. (0,0) .. controls (1.99,0.18) and (4.17,0.84) .. (6.56,1.97)   ;
\draw    (5,75.5) -- (18,75.5) ;
\draw [shift={(3,75.5)}, rotate = 0] [color={rgb, 255:red, 0; green, 0; blue, 0 }  ][line width=0.75]    (6.56,-1.97) .. controls (4.17,-0.84) and (1.99,-0.18) .. (0,0) .. controls (1.99,0.18) and (4.17,0.84) .. (6.56,1.97)   ;
\draw    (228,4.82) -- (228,16.44) ;
\draw [shift={(228,2.82)}, rotate = 90] [color={rgb, 255:red, 0; green, 0; blue, 0 }  ][line width=0.75]    (6.56,-1.97) .. controls (4.17,-0.84) and (1.99,-0.18) .. (0,0) .. controls (1.99,0.18) and (4.17,0.84) .. (6.56,1.97)   ;
\draw  [color={rgb, 255:red, 0; green, 0; blue, 0 }  ,draw opacity=0.5 ][pattern=_e442rm2w0,pattern size=11.25pt,pattern thickness=0.75pt,pattern radius=0pt, pattern color={rgb, 255:red, 0; green, 0; blue, 0}][line width=0.75]  (3,139.09) -- (103,139.09) -- (103,161.8) -- (3,161.8) -- cycle ;

\draw   (292,2) -- (307,2) -- (307,17) -- (292,17) -- cycle ;
\draw   (292,23) -- (307,23) -- (307,38) -- (292,38) -- cycle ;
\draw  [color={rgb, 255:red, 0; green, 0; blue, 0 }  ,draw opacity=0.5 ][pattern=_jhy93hey3,pattern size=3.75pt,pattern thickness=0.75pt,pattern radius=0pt, pattern color={rgb, 255:red, 0; green, 0; blue, 0}][line width=0.75]  (292,23) -- (307,23) -- (307,38) -- (292,38) -- cycle ;

\draw   (292,44) -- (307,44) -- (307,59) -- (292,59) -- cycle ;
\draw [color={rgb, 255:red, 0; green, 0; blue, 0 }  ,draw opacity=0.5 ]   (307,44) -- (292,59) ;
\draw [color={rgb, 255:red, 0; green, 0; blue, 0 }  ,draw opacity=0.5 ]   (292,44) -- (307,59) ;

\draw   (294.5,67.5) -- (304.5,67.5) -- (304.5,77.5) -- (294.5,77.5) -- cycle ;
\draw   (292,65) -- (307,65) -- (307,80) -- (292,80) -- cycle ;

\draw  [fill={rgb, 255:red, 183; green, 210; blue, 228 }  ,fill opacity=1 ] (292,107) -- (307,107) -- (307,122) -- (292,122) -- cycle ;
\draw  [fill={rgb, 255:red, 158; green, 199; blue, 215 }  ,fill opacity=1 ] (292,128) -- (307,128) -- (307,143) -- (292,143) -- cycle ;
\draw  [fill={rgb, 255:red, 191; green, 228; blue, 251 }  ,fill opacity=1 ] (292,86) -- (307,86) -- (307,101) -- (292,101) -- cycle ;
\draw    (292,155.5) -- (307,155.5) ;
\draw [shift={(307,155.5)}, rotate = 180] [color={rgb, 255:red, 0; green, 0; blue, 0 }  ][line width=0.75]    (0,3.35) -- (0,-3.35)(6.56,-1.97) .. controls (4.17,-0.84) and (1.99,-0.18) .. (0,0) .. controls (1.99,0.18) and (4.17,0.84) .. (6.56,1.97)   ;
\draw  [dash pattern={on 1.69pt off 2.76pt}][line width=1.5]  (292,170) -- (307,170) -- (307,185) -- (292,185) -- cycle ;

\draw (314,2) node [anchor=north west][inner sep=0.75pt]   [align=left] {Soft Block};
\draw (314,23) node [anchor=north west][inner sep=0.75pt]   [align=left] {Hard Block};
\draw (314,44) node [anchor=north west][inner sep=0.75pt]   [align=left] {Preplaced Block};
\draw (314,65) node [anchor=north west][inner sep=0.75pt]   [align=left] {MIB};
\draw (314,86) node [anchor=north west][inner sep=0.75pt]   [align=left] {Group 0};
\draw (313.5,107) node [anchor=north west][inner sep=0.75pt]   [align=left] {Group 1};
\draw (314,128) node [anchor=north west][inner sep=0.75pt]   [align=left] {Group 2};
\draw (314,149) node [anchor=north west][inner sep=0.75pt]   [align=left] {Boundary};
\draw (314,170) node [anchor=north west][inner sep=0.75pt]   [align=left] {Fixed-outline};
\end{tikzpicture}
}
    \caption{Example of design constraints in floorplanning.}
    \label{fig:constraint}
\vspace{-5mm}
\end{figure}

\textbf{Constraints.}
As illustrated in Figure~\ref{fig:constraint}, the floorplanning problem is subject to various constraints, which can be categorized as follows \cite{2025FloorSetChallenge,floorset}:
\begin{itemize}
    \item \textbf{Hard block constraints:} Each hard block $v_i \in \mathbf{V}_{h}$ has fixed dimensions $\bar{w}_i$ and $\bar{h}_i$.
    \item \textbf{Area constraints:} Each soft block $v_i \in \mathbf{V}_{s}$ must satisfy its target area $A_i$ constraint.
    \item \textbf{Aspect ratio constraints:} Each soft block $v_i \in \mathbf{V}_{s}$ must satisfy $\text{AR}_i^{\min} \leq w_i/h_i \leq \text{AR}_i^{\max}$.
    \item \textbf{Multi-instance block (MIB) constraints:} Blocks belonging to the same multi-instantiation group $\mathbf{M}_i \in \mathbf{M}$ must share identical dimensions.
    \item \textbf{Non-overlap constraints:} 
    The overlap area $\text{overlap}(v_i,v_j)$ between any two blocks $v_i$ and $v_j$ must be zero.
    \item \textbf{Fixed-outline constraints:} The entire floorplan must be contained within a specified fixed outline of $W_f \times H_f$.
    \item \textbf{Preplaced block constraints:} Each preplaced block $v_i \in \mathbf{V}_{p}$ has a fixed position and dimensions $(\bar{x}_i, \bar{y}_i, \bar{w}_i, \bar{h}_i)$.
    \item \textbf{Grouping constraint:} For each group $\mathbf{G}_k \in \mathbf{G}$, there must exist a spanning tree $T_k$ such that adjacent blocks in $T_k$ share a non-zero common boundary, i.e., $\text{abut}(v_i, v_j) > 0$.
    \item \textbf{Boundary constraints:} Each block in the boundary group $\mathbf{B}_j \in \{\mathbf{B}_{\text{Left}}, \mathbf{B}_{\text{Right}}, \mathbf{B}_{\text{Bottom}}, \mathbf{B}_{\text{Top}}\}$ must touch its corresponding die boundary edge $bnd_j$, indicated by $\text{touch}(v_i, bnd_j)$.
\end{itemize}

The constrained floorplanning problem is formally stated as the following optimization problem:
\begin{subequations}\label{eq:obj1}
    \begin{align}
        \min_{\boldsymbol{v}}\quad
         & \mathrm{HPWL}(\boldsymbol{v})+\mathrm{Area}(\boldsymbol{v}) \\
        \text{s.t.}\quad
         & (w_i, h_i) = (\bar{w}_i, \bar{h}_i), \quad \forall v_i\in\mathbf{V}_{h}, \label{eq:1b}\\
         & w_i h_i \ge A_i,  \quad  \forall v_i\in\mathbf{V}_{s}, \\
         & \text{AR}_i^{\min} \le w_i/h_i \le \text{AR}_i^{\max}, \quad \forall v_i\in\mathbf{V}_{s}, \\
         & (w_i, h_i) = (w_j, h_j), \quad \forall v_i, v_j \in \mathbf{M}_k, \forall\mathbf{M}_k \in \mathbf{M}, \\
         & \text{overlap}(v_i, v_j) = 0, \quad i\neq j, \forall v_i, v_j \in \mathbf{V},  \label{eq:1f}  \\
         & x_i^- \ge 0, y_i^- \ge 0, x_i^+ \le W_f, y_i^+ \le H_f, \quad \forall v_i\in\mathbf{V},  \\
         & (x_i, y_i, w_i, h_i) = (\bar{x}_i, \bar{y}_i, \bar{w}_i, \bar{h}_i), \quad \forall v_i\in\mathbf{V}_{p}, \\
         & \text{abut}(v_i, v_j) > 0, \quad \forall e_{i,j} \in T_k, \exists T_k \in \mathbf{G}_k, \mathbf{G}_k \in \mathbf{G}  \label{eq:1i} \\
         & \text{touch}(v_i, bnd_j), \quad \forall v_i \in \mathbf{B}_j, \forall \mathbf{B}_j \in \mathbf{B}.\label{eq:1j}
    \end{align}
\end{subequations}

Due to these constraints~(\ref{eq:1b})--(\ref{eq:1j}), the optimization problem in~\eqref{eq:obj1} is highly non-convex and computationally intractable to solve directly. To address this, we propose a novel three-stage analytical framework, detailed in the following section.

\begin{figure*}
    \centering
    \includegraphics[width=\textwidth]{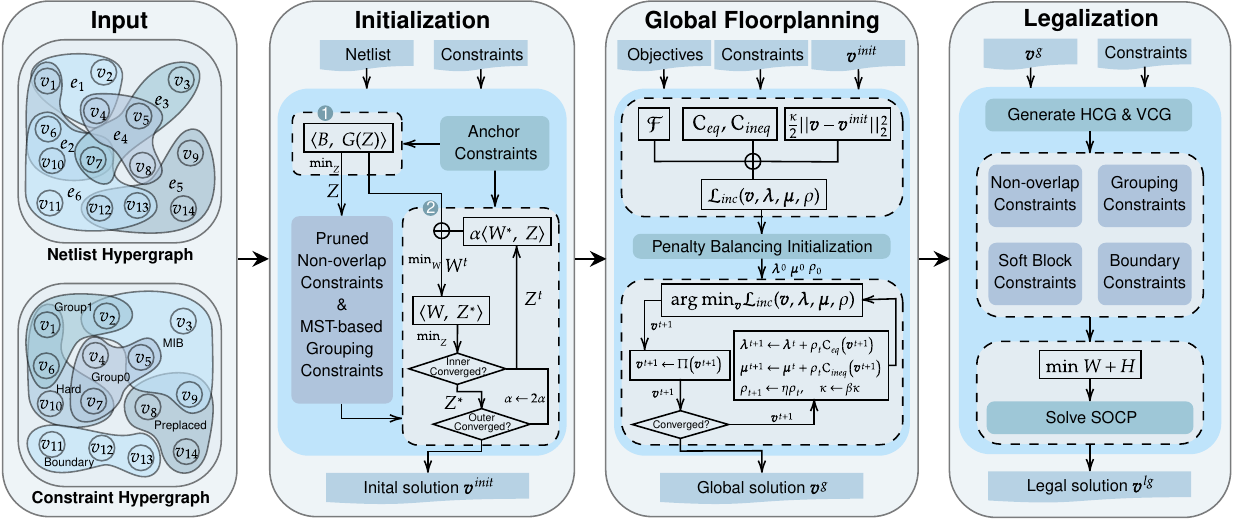}
    \caption{Overview of the three-stage constraint-driven floorplanning framework.}
    \label{fig:algorithm}
\end{figure*}
\section{Constraint-Driven Floorplanning Framework}\label{sec:alg}

Figure~\ref{fig:algorithm} shows our three-stage constraint-driven floorplanning flow. 
Initialization computes a robust seed by a two-step SDP: anchor constraints first fix global structure, then pruned non-overlap and minimum spanning tree (MST)-based grouping constraints resolve constraint violations with reduced complexity. Global floorplanning starts from this seed and applies an incremental ALM to a differentiable objective, jointly suppressing overlap, grouping, boundary, and MIB constraints via balanced penalty initialization, locality-constrained updates, and feasibility projection. Legalization transforms the near-feasible global result into a fully legal floorplan by constructing horizontal \& vertical constraint graphs and solving a unified SOCP with non-overlap, grouping, soft block, and boundary constraints. This coarse-to-fine pipeline ensures robust convergence to high-quality, strictly legal floorplans under complex coupled constraints. The following sections detail each stage of the framework.
\subsection{Initialization via Two-Step Semidefinite Programming}
High-quality initialization is critical for constrained floorplanning, where coupled constraints produce a highly non-convex landscape with many poor local minima. To efficiently obtain reliable initial solutions, we propose a two-step SDP framework that balances computational cost and geometric quality.
Unlike prior work \cite{2023SDP}, which targets global floorplanning with dense $O(n^2)$ non-overlap constraints under circular abstractions, our method focuses on robust initialization. This focus allows us to simplify the formulation by pruning unnecessary constraints and incorporating practical design requirements, while still providing a strong starting point for the subsequent rectangle-based global floorplanning stage.
\subsubsection{Basic SDP Formulation}
We adopt the SDP formulation in \cite{2023SDP}, where each block $v_i$ is abstracted as a circle centered at $\mathbf{p}_i=[x_i,y_i]^\top\in\mathbb{R}^2$ with radius $r_i = \sqrt{A_i/\pi}$.
The quadratic wirelength objective is formulated as:
\begin{equation*}
     \langle B,G\rangle = \langle A,D\rangle=\sum_{i=1}^{n}\sum_{j=1}^{n}A_{ij}\|\mathbf{p}_i-\mathbf{p}_j\|_2^2.
\end{equation*}

Here, $A\in\mathbb{R}^{n\times n}$ is the connectivity matrix constructed from the nets $\mathbf{E}$, and $D\in\mathbb{R}^{n\times n}$ is the squared Euclidean distance matrix with $D_{ij}=\|\mathbf{p}_i-\mathbf{p}_j\|_2^2$. 
Correspondingly, $G=XX^\top\in\mathbb{S}_+^n$ is the Gram matrix constructed by the coordinate matrix $X=[\mathbf{p}_1,\dots,\mathbf{p}_n]^\top\in\mathbb{R}^{n\times 2}$, and $B = \mathrm{diag}(A\mathbf{1} + A^\top \mathbf{1}) - 2A$ is the Laplacian-like matrix derived from $A$.
To obtain a convex relaxation, \cite{2023SDP} introduces the lifted matrix:
\begin{equation*}
    Z=
    \begin{bmatrix}
        I_2 & X^\top \\
        X & G
    \end{bmatrix}\in\mathbb{S}_+^{n+2},
\end{equation*}
where $I_2$ denotes the $2 \times 2$ identity matrix. By the Schur complement, the non-convex constraint $G=XX^\top$ is relaxed to $Z\succeq0$. The remaining non-convexity is handled by rank control in the iterative refinement stage.

\subsubsection{Step I: Trial Placement with Anchor Constraints}
To establish a global position, we first solve a relaxed SDP that includes only wirelength and fixed-anchor constraints, while deferring complex overlap constraints.

\paragraph{Anchor Constraints.}
We unify preplaced and boundary blocks as a set of anchor nodes, whose coordinates are fixed a priori. Let $\mathcal{C}_\text{anchor} = \mathbf{V}_p \cup \mathbf{B}$ denote the set of anchor blocks.
For each anchor block $v_i \in \mathcal{C}_\text{anchor}$, we enforce the following linear equality constraints:
\begin{equation}
\begin{gathered}
x_i = \bar{x}_i,\; y_i = \bar{y}_i,\quad \forall v_i \in \mathbf{V}_p, \\[2pt]
    \begin{aligned}
         & x_i^- = x_{\min}, \quad   \forall v_i \in \mathbf{B}_{\text{Left}},
         & x_i^+ = x_{\max},  \quad   \forall v_i \in \mathbf{B}_{\text{Right}},  \\
         & y_i^- = y_{\min}, \quad    \forall v_i \in \mathbf{B}_{\text{Bottom}},
         & y_i^+ = y_{\max},  \quad   \forall v_i \in \mathbf{B}_{\text{Top}}.
    \end{aligned}
\end{gathered}
\label{eq:anchor_constraints}
\end{equation}

\input{include/Figure/init.tex}

\paragraph{First Step SDP Problem.}
Therefore, the first-step SDP problem is formulated as:
\begin{equation*}
    \begin{aligned}
        \min_{Z} \quad & \langle B, G(Z) \rangle                           \\
        \text{s.t.} \quad
                       & Z \succeq 0,                                      \\
                       & \text{Constraints \eqref{eq:anchor_constraints}}.
    \end{aligned}
\end{equation*}

This solution captures the global structure and respects fixed preplaced and boundary locations, providing a strong foundation for the next step.

\subsubsection{Step II: Constraint-Driven SDP Initialization} 
As illustrated in Figure \ref{fig:init}, based on the Step I solution, we construct constraints to refine the solution, improving geometric quality and preparing for subsequent global floorplanning.

\paragraph{Pruned Non-overlap Constraints}
Instead of enforcing all pairwise non-overlap constraints, we construct a candidate set $\mathcal{C}_\text{overlap}$ based on relative positions. From the Step I solution $Z^{(1)}$, we identify overlapping pairs using the following rule:
\begin{equation*}
    \mathcal{C}_\text{overlap} = \{(i,j) \mid \|\mathbf{p}_i^{(1)} - \mathbf{p}_j^{(1)}\|_2^2 < (r_i + r_j)^2\}.
\end{equation*}

We enforce non-overlap constraints only for pairs in $\mathcal{C}_\text{overlap}$:
\begin{equation}
    \|\mathbf{p}_i - \mathbf{p}_j\|_2^2 \ge (r_i + r_j)^2 \quad \forall (i,j) \in \mathcal{C}_\text{overlap}.
    \label{eq:non_overlap1}
\end{equation}

\paragraph{Grouping Constraints via Minimum Spanning Tree}
For each predefined group, we construct a minimum spanning tree based on the Step I solution. 
Let $\text{MST}(\mathbf{G}_k)$ denote the set of MST edges for each group $\mathbf{G}_k$.
For each edge $(i,j) \in \text{MST}(\mathbf{G}_k)$, the distance upper bound is defined as:
\begin{equation*}
\begin{aligned}
    &\phi_i = \max\big(\mathrm{AR}_i^{\min} + (\mathrm{AR}_i^{\min})^{-1},\;
                 \mathrm{AR}_i^{\max} + (\mathrm{AR}_i^{\max})^{-1}\big), \\
    &\psi_{ij} = \frac{1}{2}\left(\sqrt{A_i \phi_i} + \sqrt{A_j \phi_j}\right).
\end{aligned}
\end{equation*}

Then we enforce the following constraint:
\begin{equation}
    \|\mathbf{p}_i - \mathbf{p}_j\|_2^2 \leq \psi_{ij}^2, \quad \forall (i,j) \in \text{MST}(\mathbf{G}_k), \forall \mathbf{G}_k \in \mathbf{G}.
    \label{eq:grouping_constraints}
\end{equation}

\paragraph{Second-Step SDP Problem.}
To ensure the relaxed solution maps accurately to a 2D plane, the matrix $Z$ must satisfy $\text{rank}(Z) = 2$. Following \cite{2023SDP}, we convexify this constraint using a direction matrix $W \in \mathbb{S}_+^{n+2}$. The second-stage SDP problem is formulated as:
\begin{equation*}
    \begin{aligned}
        \min_{Z} \quad & \langle B, G(Z) \rangle + \delta  \langle W, Z \rangle                                                       \\
        \text{s.t.} \quad
                       & Z \succeq 0,                                                                                                \\
                       & \text{Constraints \eqref{eq:anchor_constraints}, \eqref{eq:non_overlap1}, \eqref{eq:grouping_constraints}}.
    \end{aligned}
\end{equation*}

Here, $\delta$ is the penalty coefficient for rank control. We follow \cite{2023SDP} to solve this problem, which alternately optimizes $W$ and $Z$ in an inner loop, while increasing $\delta$ in an outer loop, until an approximate rank-2 solution $Z^*$ is obtained. Finally, the initial solution $\boldsymbol{v}^{init}$ is extracted from $Z^*$ for the next global floorplanning stage.

\subsection{Global Floorplanning via Multi-Constraint Augmented Lagrangian Method}
The initialization stage provides a reasonable coarse topology that serves as a good starting point for global floorplanning. Building upon this initial solution, we formulate global floorplanning as a differentiable constrained optimization problem that jointly minimizes wirelength and bounding-box area, while modeling overlap, grouping, boundary, and multi-instantiation block constraints with differentiable surrogates. We solve it using an initial-solution-aware incremental ALM, which progressively tightens constraints and improves optimization stability.

\subsubsection{Floorplan Area}
To make the bounding box differentiable, we use the log-sum-exp (LSE) function with smoothing coefficient $\gamma$ to approximate the extrema of module coordinates. The smoothed extrema in the $x$-direction are formulated as:
\begin{equation*}
    \begin{aligned}
    &\hat{x}_{\min}(\boldsymbol{v}):=-\gamma\ln\sum_{i=1}^{n}\exp\left(-\frac{x_i^-}{\gamma}\right),\\
    &\hat{x}_{\max}(\boldsymbol{v}):=\gamma\ln\sum_{i=1}^{n}\exp\left(\frac{x_i^+}{\gamma}\right),
    \end{aligned}
\end{equation*}
and the $y$-direction terms are defined analogously. Therefore, we obtain a fully differentiable surrogate for the bounding-box area $\mathcal{A}(\boldsymbol{v}) $ as follows:
\begin{equation*}
    \mathcal{A}(\boldsymbol{v}) := \left(\hat{x}_{\max}(\boldsymbol{v})-\hat{x}_{\min}(\boldsymbol{v})\right)
    \left(\hat{y}_{\max}(\boldsymbol{v})-\hat{y}_{\min}(\boldsymbol{v})\right).
\end{equation*}

\subsubsection{Multi-Instantiation Block Constraints}
For each MIB group $\mathbf{M}_i\in\mathbf{M}$, we enforce shape consistency through identical aspect ratios. We first compute the group-wise mean aspect ratio, denoted as $\overline{ar}_i(\boldsymbol{v})$:
\begin{equation*}
    \overline{ar}_i(\boldsymbol{v}):=\frac{1}{|\mathbf{M}_i|}\sum_{v_j\in\mathbf{M}_i}\frac{w_j}{h_j}.
\end{equation*}

Using this mean, we define the inconsistency penalty $\mathcal{M}_i(\boldsymbol{v})$ for group $\mathbf{M}_i$:
\begin{equation*}
    \mathcal{M}_i(\boldsymbol{v}):=\frac{1}{|\mathbf{M}_i|}\sum_{v_j\in\mathbf{M}_i}
    \left(\frac{w_j}{h_j}-\overline{ar}_i(\boldsymbol{v})\right)^2.
\end{equation*}

This variance-like term quantifies inconsistency among instances in the same MIB group; $\mathcal{M}_i(\boldsymbol{v})=0$ indicates perfect shape consistency.
Finally, we stack all group-wise penalties into a vector:
\begin{equation*}
    \mathcal{M}(\boldsymbol{v}):=\left[\mathcal{M}_1(\boldsymbol{v}),\mathcal{M}_2(\boldsymbol{v}),\dots,\mathcal{M}_{|\mathbf{M}|}(\boldsymbol{v})\right]^\top.
\end{equation*}

\begin{figure}[htbp]

\tikzset{every picture/.style={line width=0.75pt}} %

\begin{tikzpicture}[x=0.75pt,y=0.75pt,yscale=-0.75,xscale=0.75]

\draw  [fill={rgb, 255:red, 227; green, 236; blue, 242 }  ,fill opacity=0.8 ] (45,75) -- (145,75) -- (145,175) -- (45,175) -- cycle ;
\draw  [fill={rgb, 255:red, 227; green, 236; blue, 242 }  ,fill opacity=0.8 ] (105,135) -- (205,135) -- (205,235) -- (105,235) -- cycle ;
\draw  [fill={rgb, 255:red, 183; green, 210; blue, 228 }  ,fill opacity=0.8 ] (105,135) -- (145,135) -- (145,175) -- (105,175) -- cycle ;
\draw    (95,114) -- (155,114) ;
\draw [shift={(155,114)}, rotate = 180] [color={rgb, 255:red, 0; green, 0; blue, 0 }  ][line width=0.75]    (0,3.35) -- (0,-3.35)   ;
\draw [shift={(95,114)}, rotate = 180] [color={rgb, 255:red, 0; green, 0; blue, 0 }  ][line width=0.75]    (0,3.35) -- (0,-3.35)   ;
\draw    (164,125) -- (164,185) ;
\draw [shift={(164,185)}, rotate = 270] [color={rgb, 255:red, 0; green, 0; blue, 0 }  ][line width=0.75]    (0,3.35) -- (0,-3.35)   ;
\draw [shift={(164,125)}, rotate = 270] [color={rgb, 255:red, 0; green, 0; blue, 0 }  ][line width=0.75]    (0,3.35) -- (0,-3.35)   ;
\draw    (108,181) -- (142,181) ;
\draw [shift={(142,181)}, rotate = 180] [color={rgb, 255:red, 0; green, 0; blue, 0 }  ][line width=0.75]    (0,2.24) -- (0,-2.24)   ;
\draw [shift={(108,181)}, rotate = 180] [color={rgb, 255:red, 0; green, 0; blue, 0 }  ][line width=0.75]    (0,2.24) -- (0,-2.24)   ;
\draw    (98,138) -- (98,172) ;
\draw [shift={(98,172)}, rotate = 270] [color={rgb, 255:red, 0; green, 0; blue, 0 }  ][line width=0.75]    (0,2.24) -- (0,-2.24)   ;
\draw [shift={(98,138)}, rotate = 270] [color={rgb, 255:red, 0; green, 0; blue, 0 }  ][line width=0.75]    (0,2.24) -- (0,-2.24)   ;
\draw  [fill={rgb, 255:red, 0; green, 0; blue, 0 }  ,fill opacity=1 ] (94,126) .. controls (94,125.45) and (94.45,125) .. (95,125) .. controls (95.55,125) and (96,125.45) .. (96,126) .. controls (96,126.55) and (95.55,127) .. (95,127) .. controls (94.45,127) and (94,126.55) .. (94,126) -- cycle ;
\draw  [fill={rgb, 255:red, 0; green, 0; blue, 0 }  ,fill opacity=1 ] (153,185) .. controls (153,184.45) and (153.45,184) .. (154,184) .. controls (154.55,184) and (155,184.45) .. (155,185) .. controls (155,185.55) and (154.55,186) .. (154,186) .. controls (153.45,186) and (153,185.55) .. (153,185) -- cycle ;
\draw (345,160) node  {\includegraphics[width=0.58\hsize]{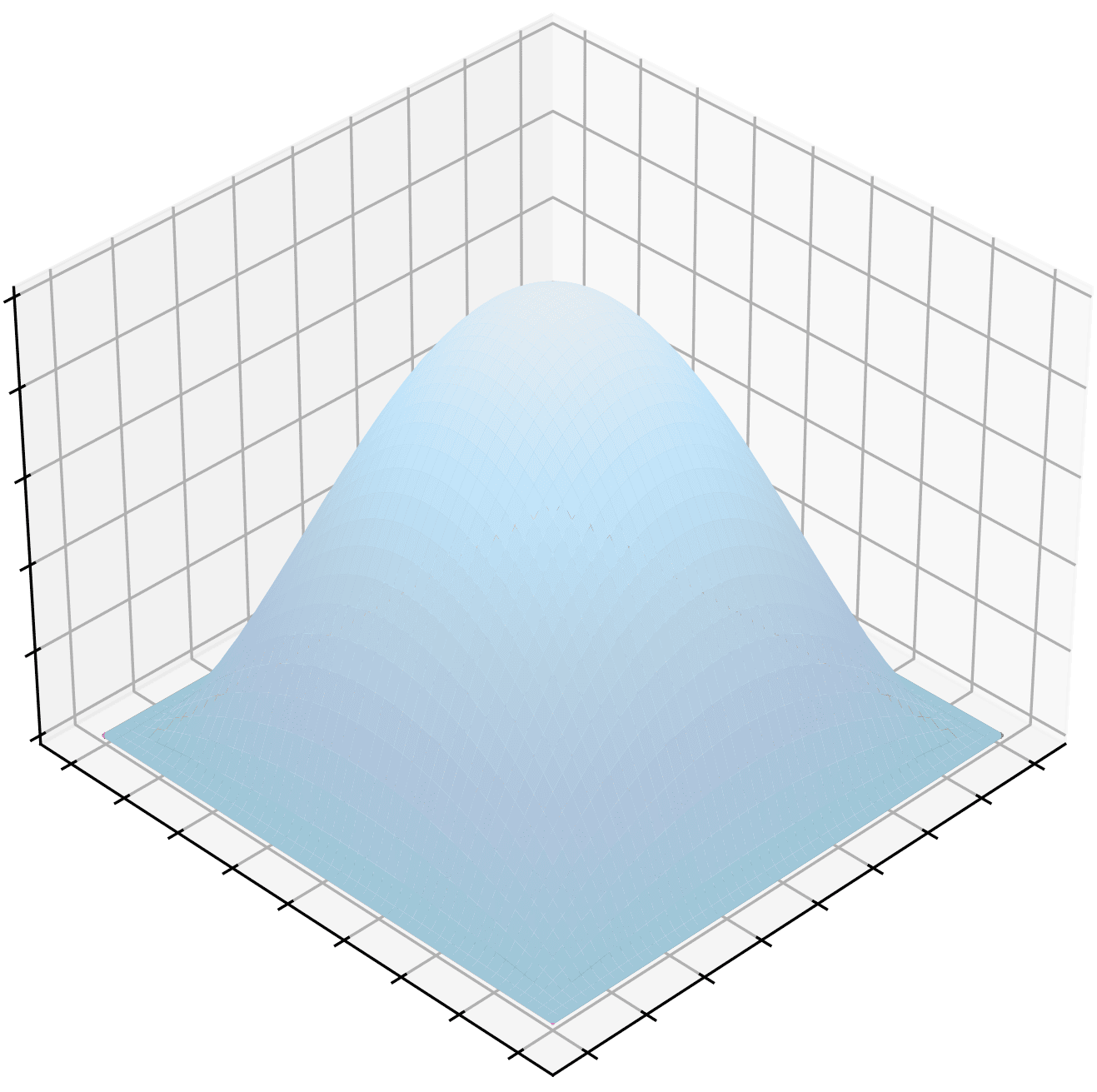}};

\draw (110,85) node [anchor=north west][inner sep=0.75pt]   [align=left] {$d_{ij}^{x}$};
\draw (170,142) node [anchor=north west][inner sep=0.75pt]   [align=left] {$d_{ij}^{y}$};
\draw (115,185) node [anchor=north west][inner sep=0.75pt]  [font=\normalsize] [align=left] {$o_{ij}^{x}$};
\draw (75,140) node [anchor=north west][inner sep=0.75pt]  [font=\normalsize] [align=left] {$o_{ij}^{y}$};
\draw (112,147) node [anchor=north west][inner sep=0.75pt]  [font=\normalsize] [align=left] {$O_{ij}$};
\draw (77,110) node [anchor=north west][inner sep=0.75pt]  [font=\normalsize] [align=left] {$v_{i}$};
\draw (152,188) node [anchor=north west][inner sep=0.75pt]  [font=\normalsize] [align=left] {$v_{j}$};
\draw (325,75) node [anchor=north west][inner sep=0.75pt]  [font=\normalsize] [align=left] {$O_{ij}(\boldsymbol{v})$};
\draw (385,260) node [anchor=north west][rotate=-325][inner sep=0.75pt]   [align=left] {$x_i - x_j$};
\draw (265,230) node [anchor=north west][rotate=-35][inner sep=0.75pt]   [align=left] {$y_i - y_j$};
\end{tikzpicture}
\caption{Bell-shaped pairwise overlap model.}
\label{fig:overlap}
\vspace{-1mm}
\end{figure}

\subsubsection{Overlap Constraint}
To model non-overlap with smooth penalties, we use a bell-shaped function~\cite{aplace} to smooth the pairwise overlap constraints as illustrated in Figure~\ref{fig:overlap}. For each pair $(i,j)$, we first define the geometric quantities:
\begin{equation*}
\zeta_{ij}^x(\boldsymbol{v}):=\frac{w_i+w_j}{2},\quad
    \vartheta_{ij}^x(\boldsymbol{v}):=\frac{|w_i-w_j|}{2},\quad
    d_{ij}^x(\boldsymbol{v}):=|x_i-x_j|,
\end{equation*}
where $\zeta_{ij}^x$ is the contact threshold along $x$, $\vartheta_{ij}^x$ characterizes size mismatch, and $d_{ij}^x$ is center distance.
Based on these quantities, the $x$-direction overlap term $O_{ij}^x(\boldsymbol{v})$ is defined as:
\begin{equation*}
    O_{ij}^x(\boldsymbol{v}):=
    \begin{cases}
        \zeta_{ij}^x(\boldsymbol{v})\!\left(1-\dfrac{(d_{ij}^x(\boldsymbol{v}))^2}{\zeta_{ij}^x(\boldsymbol{v})\vartheta_{ij}^x(\boldsymbol{v})}\right), & 0\le d_{ij}^x(\boldsymbol{v}) < \vartheta_{ij}^x(\boldsymbol{v}), \\
        \dfrac{(d_{ij}^x(\boldsymbol{v})-\zeta_{ij}^x(\boldsymbol{v}))^2}{\zeta_{ij}^x(\boldsymbol{v})-\vartheta_{ij}^x(\boldsymbol{v})}, & \vartheta_{ij}^x(\boldsymbol{v})\le d_{ij}^x(\boldsymbol{v})\le \zeta_{ij}^x(\boldsymbol{v}), \\
        0, & \zeta_{ij}^x(\boldsymbol{v})<d_{ij}^x(\boldsymbol{v}).
    \end{cases}
\end{equation*}

The corresponding $y$-direction overlap term $O_{ij}^y(\boldsymbol{v})$ is defined analogously.
These piecewise functions are
continuously differentiable in practical ranges and decays to zero once modules are sufficiently separated. 
To capture the 2D overlap, the pairwise overlap $O_{ij}(\boldsymbol{v})$ is formulated as the product of the 1D terms:
\begin{equation*}
    O_{ij}(\boldsymbol{v}):=O_{ij}^x(\boldsymbol{v})\cdot O_{ij}^y(\boldsymbol{v}).
\end{equation*}

The multiplicative form ensures that overlap penalty is significant only when projections overlap in both axes.
The normalized overlap violation $\mathcal{O}_i(\boldsymbol{v})$ for block $v_i$ is:
\begin{equation*}
\mathcal{O}_i(\boldsymbol{v}):=\frac{\sum_{j\neq i}O_{ij}(\boldsymbol{v})}{2A_i},
\end{equation*}
where normalization by area $A_i$ avoids bias toward large modules.
The block-wise overlap penalties are then stacked into a vector:
\begin{equation*}
    \mathcal{O}(\boldsymbol{v}):=\left[\mathcal{O}_1(\boldsymbol{v}),\mathcal{O}_2(\boldsymbol{v}),\dots,\mathcal{O}_n(\boldsymbol{v})\right]^\top.
\end{equation*}

\subsubsection{Grouping Constraints}
For each group $\mathbf{G}_i\in\mathbf{G}$, we build a complete graph and define edge weights using the abutment potential. We first define directional gap overflows between block pairs:
\begin{equation*}
\begin{aligned}
&ad_{jk}^x(\boldsymbol{v}):=\max\!\left(0,\;|x_j-x_k|+\ell-\frac{w_j+w_k}{2}\right),\\
&ad_{jk}^y(\boldsymbol{v}):=\max\!\left(0,\;|y_j-y_k|+\ell-\frac{h_j+h_k}{2}\right), 
\end{aligned}
\end{equation*}
where $\ell$ is a small tolerance parameter that allows for slight gaps.
Using these directional terms, we define the edge-wise penalty as:
\begin{equation*}
\psi_{jk}(\boldsymbol{v}) := \left(ad_{jk}^x(\boldsymbol{v})\right)^2 + \left(ad_{jk}^y(\boldsymbol{v})\right)^2,
\end{equation*}
which provides a smooth isotropic penalty and emphasizes large abutment deviations.
To mitigate undesired interactions among constraints, we consider only edges in $\mathrm{MST}(\mathbf{G}_i)$ and define the group-level penalty as:
\begin{equation*} \mathcal{G}_i(\boldsymbol{v}):=\sum_{(j,k)\in\mathrm{MST}(\mathbf{G}_i)}\psi_{jk}(\boldsymbol{v}), \end{equation*}

Using only MST edges preserves group connectivity while limiting constraint coupling.
Finally, we stack all group penalties into a compact vector:
\begin{equation*}
    \mathcal{G}(\boldsymbol{v}):=\left[\mathcal{G}_1(\boldsymbol{v}),\mathcal{G}_2(\boldsymbol{v}),\dots,\mathcal{G}_{|\mathbf{G}|}(\boldsymbol{v})\right]^\top.
\end{equation*}

\subsubsection{Boundary Constraint}
For boundary-constrained blocks, we define normalized boundary overflow using the smoothed box coordinates. We first give the $x$-direction terms:
\begin{equation*}
    \begin{aligned}  
    &\mathcal{B}_{\mathrm{Left}}(\boldsymbol{v}):=
    \sum_{i\in\mathbf{B}_{\mathrm{Left}}}
    \frac{x_i^- -\hat{x}_{\min}(\boldsymbol{v})}{\hat{x}_{\max}(\boldsymbol{v})-\hat{x}_{\min}(\boldsymbol{v})},\\
    &\mathcal{B}_{\mathrm{Right}}(\boldsymbol{v}):=
    \sum_{i\in\mathbf{B}_{\mathrm{Right}}}
    \frac{\hat{x}_{\max}(\boldsymbol{v})-x_i^+}{\hat{x}_{\max}(\boldsymbol{v})-\hat{x}_{\min}(\boldsymbol{v})},
    \end{aligned}
\end{equation*}
and the $y$-direction terms are defined analogously.
Each term measures normalized boundary slack for one side and becomes zero when constrained modules exactly touch that side. Normalization by current box span reduces scale sensitivity across different floorplan sizes.
The boundary constraint vector is:
\begin{equation*}
    \mathcal{B}(\boldsymbol{v}):=
    \left[\mathcal{B}_{\mathrm{Left}}(\boldsymbol{v}),\mathcal{B}_{\mathrm{Right}}(\boldsymbol{v}),\mathcal{B}_{\mathrm{Top}}(\boldsymbol{v}),\mathcal{B}_{\mathrm{Bottom}}(\boldsymbol{v})\right]^\top.
\end{equation*}

\subsubsection{Incremental Augmented Lagrangian Optimization}
Using the above definitions, we write the constraints in compact form:
\begin{equation*}
    \mathcal{C}_{\mathrm{eq}}(\boldsymbol{v}):=
    \begin{bmatrix}
        \mathcal{M}(\boldsymbol{v}) \\
        \mathcal{G}(\boldsymbol{v}) \\
        \mathcal{B}(\boldsymbol{v})
    \end{bmatrix},
    \qquad
    \mathcal{C}_{\mathrm{ineq}}(\boldsymbol{v}):=\mathcal{O}(\boldsymbol{v}).
\end{equation*}

Equality constraints include MIB, grouping, and boundary consistency, while overlap is treated as an inequality with penalty augmentation. The global floorplanning problem is thus:
\begin{equation*}
    \begin{aligned}
        \min_{\boldsymbol{v}}\quad & \mathcal{F}(\boldsymbol{v}) = \mathcal{W}(\boldsymbol{v}) + \alpha \mathcal{A}(\boldsymbol{v}) \\
        \text{s.t.}\quad & \mathcal{C}_{\mathrm{eq}}(\boldsymbol{v})=\mathbf{0}, \\
        & \mathcal{C}_{\mathrm{ineq}}(\boldsymbol{v})\le \mathbf{0},
    \end{aligned}
\end{equation*}
where $\mathcal{W}(\cdot)$ is the smoothed HPWL model~\cite{hsu2013tsv}, $\alpha = \tfrac{\mathcal{W}(\boldsymbol{v}^{init})}{\mathcal{A}(\boldsymbol{v}^{init})}$ is a scaling factor to balance wirelength and area, and $\boldsymbol{v}^{init}$ is the initial solution from the initialization stage.

For equality multipliers $\boldsymbol{\lambda}$, inequality multipliers $\boldsymbol{\mu}\ge\mathbf{0}$, and penalty parameter $\rho>0$, the standard augmented Lagrangian is:
\begin{equation*}
    \begin{aligned}
        \mathcal{L}_{\rho}(\boldsymbol{v},\boldsymbol{\lambda},\boldsymbol{\mu})
         & := \mathcal{F}(\boldsymbol{v})+\boldsymbol{\lambda}^\top\mathcal{C}_{\mathrm{eq}}(\boldsymbol{v})
        +\frac{\rho}{2}\left\|\mathcal{C}_{\mathrm{eq}}(\boldsymbol{v})\right\|_2^2 \\
         & \qquad \qquad \qquad +\boldsymbol{\mu}^\top\mathcal{C}_{\mathrm{ineq}}(\boldsymbol{v})
        +\frac{\rho}{2}\left\|\left[\mathcal{C}_{\mathrm{ineq}}(\boldsymbol{v})\right]_+\right\|_2^2,
    \end{aligned}
\end{equation*}
where $[\cdot]_+=\max(\cdot,0)$ is applied element-wise.

Because the initial solution $\boldsymbol{v}^{init}$ already has a good global topology, directly applying the standard ALM can cause severe topological disruption due to unconstrained step sizes and unbalanced initial penalties.
To mitigate this effect, we introduce the following incremental strategy to gradually tighten constraints and guide the solution toward feasibility while preserving the initial topology. 

\paragraph{Locality Constraining}
To prevent large disruptive steps in early iterations, we add a proximal term to the augmented Lagrangian:
\begin{equation*}
    \mathcal{L}_{\mathrm{inc}}(\boldsymbol{v}, \boldsymbol{\lambda}, \boldsymbol{\mu}, \rho) = \mathcal{L}_{\rho}(\boldsymbol{v}, \boldsymbol{\lambda}, \boldsymbol{\mu}) + \frac{\kappa}{2}\|\boldsymbol{v} - \boldsymbol{v}^{init}\|_2^2,
\end{equation*}
where $\kappa > 0$ is a dynamically decaying weight. This term restricts early-step magnitudes, allowing the solver to resolve constraint violations incrementally while preserving macroscopic placement topology.

\paragraph{Penalty Balancing}
To prevent the initial constraint penalty from dominating the objective and abruptly disrupting the floorplan, we balance the initial penalty parameter $\rho_0$. Specifically, we compute $\rho_0$ by matching the infinity norms of the objective gradient and the maximum constraint violation at the initial solution $\boldsymbol{v}^{init}$:
\begin{equation*}
    \rho_0 = \frac{\|\nabla \mathcal{F}(\boldsymbol{v}^{init})\|_{\infty}}{\max(1, \|\mathcal{C}(\boldsymbol{v}^{init})\|_{\infty})},
\end{equation*}
where $\mathcal{C}(\cdot) = \mathcal{C}_{\mathrm{eq}}(\cdot) \cup \left[\mathcal{C}_{\mathrm{ineq}}(\cdot)\right]_+$ includes both equality and inequality violations. Instead of initializing multipliers to zero, which may require solving a highly ill-conditioned least-squares problem for KKT stationarity, we use $\rho_0$ to construct an approximate projection for the initial multipliers:
\begin{equation*}
    \boldsymbol{\lambda}^0 = \rho_0 \mathcal{C}_{\mathrm{eq}}(\boldsymbol{v}^{init}), \quad \boldsymbol{\mu}^0 = \rho_0 \left[\mathcal{C}_{\mathrm{ineq}}(\boldsymbol{v}^{init})\right]_+.
\end{equation*}

This initialization provides a high-quality estimate that aligns the Lagrangian gradient near zero at $\boldsymbol{v}^{init}$.

\paragraph{Handling Remaining Constraints}
For the remaining constraints, we enforce them explicitly through projection and gradient masking. 
During the variable update phase, the fixed-outline, area and aspect-ratio constraints are enforced by direct projection:
\begin{equation*}
\boldsymbol{v}^{t+1}\leftarrow\Pi(\boldsymbol{v}^{t+1}),
\end{equation*}
where $\Pi(\cdot)$ is the projection onto the fixed-outline and aspect-ratio feasible sets.
During the gradient evaluation phase, we apply a binary mask $\mathbf{m}$ to freeze the dimensions of the hard blocks and all variables of the preplaced blocks:
\begin{equation*}
\nabla_{\boldsymbol{v}}\mathcal{L}_{\mathrm{inc}}\leftarrow\mathbf{m}\odot\nabla_{\boldsymbol{v}}\mathcal{L}_{\mathrm{inc}},
\end{equation*}
\paragraph{Primal and Dual Updates}
In outer iteration $t$, we first approximately solve the primal subproblem:
\begin{equation*}
    \boldsymbol{v}^{t+1}\approx
    \arg\min_{\boldsymbol{v}}\mathcal{L}_{\mathrm{inc}}(\boldsymbol{v},\boldsymbol{\lambda}^t,\boldsymbol{\mu}^t, \rho_t).
\end{equation*}

Dual variables are then updated via standard ascent and non-negative orthant projection to maintain KKT complementarity:
\begin{equation*}
    \boldsymbol{\lambda}^{t+1}=\boldsymbol{\lambda}^{t}+\rho_t\,\mathcal{C}_{\mathrm{eq}}(\boldsymbol{v}^{t+1}), \quad \boldsymbol{\mu}^{t+1}=\left[\boldsymbol{\mu}^{t}+\rho_t\,\mathcal{C}_{\mathrm{ineq}}(\boldsymbol{v}^{t+1})\right]_+.
\end{equation*}

Finally, the global floorplanning workflow is summarized in Algorithm \ref{alg:inc_alm}. This workflow jointly optimizes wirelength and area while resolving multiple constraints to yield a near-legal solution.

\begin{algorithm}[htbp]
\caption{Incremental Augmented Lagrangian Optimization}
\label{alg:inc_alm}
\begin{algorithmic}[1]
\REQUIRE Initial solution $\boldsymbol{v}^{init}$, target constraint tolerance $\varepsilon_{\mathrm{target}}$, maximum iterations $T$.
\ENSURE Global solution $\boldsymbol{v}^{g}$.
\STATE \textbf{// Step 1: Penalty Balancing Initialization}
\STATE $\boldsymbol{g}_f \leftarrow \nabla \mathcal{F}(\boldsymbol{v}^{init})$
\STATE $\boldsymbol{c}_{\mathrm{eq}} \leftarrow \mathcal{C}_{\mathrm{eq}}(\boldsymbol{v}^{init}), \quad \boldsymbol{c}_{\mathrm{ineq}} \leftarrow \mathcal{C}_{\mathrm{ineq}}(\boldsymbol{v}^{init})$
\STATE $c_{\max} \leftarrow \max\big(1, \|\boldsymbol{c}_{\mathrm{eq}}\|_{\infty}, \|[\boldsymbol{c}_{\mathrm{ineq}}]_+\|_{\infty}\big)$
\STATE Balance initial penalty: $\rho_0 \leftarrow \|\boldsymbol{g}_f\|_{\infty} / c_{\max}$
\STATE Initialize multipliers: $\boldsymbol{\lambda}^0 \leftarrow \rho_0 \boldsymbol{c}_{\mathrm{eq}}, \quad \boldsymbol{\mu}^0 \leftarrow \rho_0 [\boldsymbol{c}_{\mathrm{ineq}}]_+$
\STATE Initialize locality weight $\kappa_0 > 0$ and decay factor $\beta \in (0, 1)$
\STATE \textbf{// Step 2: Incremental Optimization Loop}
\FOR{$t = 0, 1, \dots, T-1$}
    \STATE $\displaystyle \boldsymbol{v}^{t+1} \leftarrow \arg\min_{\boldsymbol{v}} \mathcal{L}_{\mathrm{inc}}(\boldsymbol{v}, \boldsymbol{\lambda}^t, \boldsymbol{\mu}^t, \rho_t)$
    \STATE Apply feasibility projection: $\boldsymbol{v}^{t+1}\leftarrow\Pi(\boldsymbol{v}^{t+1})$
    \IF{$\left\|\mathcal{C}(\boldsymbol{v}^{t+1})\right\|_{\infty} \le \varepsilon_{\mathrm{target}}$}
        \STATE \textbf{break} \COMMENT{Convergence achieved}
    \ENDIF
    \STATE $\boldsymbol{\lambda}^{t+1} \leftarrow \boldsymbol{\lambda}^t + \rho_t \mathcal{C}_{\mathrm{eq}}(\boldsymbol{v}^{t+1})$
    \STATE $\boldsymbol{\mu}^{t+1} \leftarrow \left[\boldsymbol{\mu}^t + \rho_t \mathcal{C}_{\mathrm{ineq}}(\boldsymbol{v}^{t+1})\right]_+$
    \STATE Update penalty $\rho_{t+1} \leftarrow \eta \rho_t$
    \STATE Relax locality constraint: $\kappa_{t+1} \leftarrow \beta \kappa_t$
\ENDFOR
\RETURN $\boldsymbol{v}^{g} \leftarrow \boldsymbol{v}^{t+1}$
\end{algorithmic}
\end{algorithm}

\subsection{Legalization via Second-Order Cone Programming}
\label{subsec:socp_legalization}
Although the global solution $\boldsymbol{v}^{g}$ establishes a high-quality result, it may still contain residual overlaps and minor constraint violations. To obtain a fully legal floorplan that satisfies all constraints, we formulate the legalization process as a SOCP problem. This formulation enables the simultaneous enforcement of all constraints while further minimizing the bounding box area.

\subsubsection{Constraint Graph and Non-Overlap Constraints}

From the global floorplanning solution, we construct a horizontal constraint graph $G_h = (\mathcal{V}, \mathcal{E}_h)$ and a vertical constraint graph $G_v = (\mathcal{V}, \mathcal{E}_v)$ via method from~\cite{2023rotationfp}. For any edge in $G_h$ and $G_v$, we enforce non-overlap constraints to ensure legal placement:
\begin{equation}
    \begin{cases}
        x_i^+ \le x_j^-, & \forall (i,j) \in \mathcal{E}_h, \\
        y_i^+ \le y_j^-, & \forall (i,j) \in \mathcal{E}_v.
    \end{cases}
\label{eq:non_overlap}
\end{equation}

\subsubsection{Soft Block Constraints}
For soft blocks $v_i \in \mathbf{V}_s$ with target area $A_i$ constraint and aspect ratio limits $\text{AR}_i^{\min} \le w_i/h_i \le \text{AR}_i^{\max}$, the geometric requirements are reformulated into a combination of a second-order cone constraint and linear inequalities:
\begin{equation}
 \left\| \begin{pmatrix} h_i - w_i \\ 2\sqrt{A_i} \end{pmatrix} \right\|_2 \le h_i + w_i, \quad \text{AR}_i^{\min}h_i \leq w_i \leq \text{AR}_i^{\max}h_i.\\
 \quad \forall v_i \in \mathbf{V}_s.  \label{eq:lg:soft_block}
\end{equation}

This convex formulation perfectly models both the lower bound on the block area and the limits on its shape. Consequently, it allows the optimizer to dynamically adjust block dimensions while maintaining mathematical tractability within an SOCP-based approach.

\begin{figure}
    \centering
    \includegraphics[width=0.82\hsize]{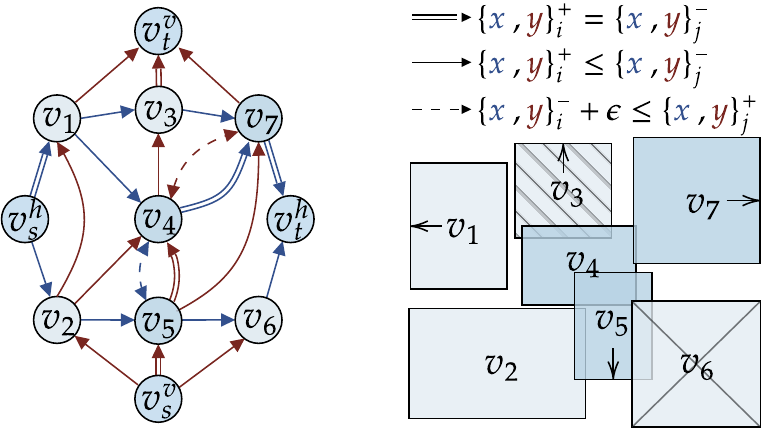}
    \caption{Global solution and constraint graph with design constraints.}
    \label{fig:cg}
\end{figure}

\subsubsection{Grouping Constraints}
To ensure physical abutment while maintaining feasibility, we enforce zero-gap abutment in one direction and a minimum overlap in the orthogonal direction. For each edge $(i,j)$ in the MST of group $\mathbf{G}_k \in \mathbf{G}$, the primary abutment direction is determined by the global solution. The constraints are defined as:
\begin{equation}
\begin{cases}
x_i^+ = x_j^-, \quad
y_i^- + \epsilon \le y_j^+,\quad
y_j^- + \epsilon \le y_i^+, & \text{if } (i,j) \in \mathcal{E}_h, \\
y_i^+ = y_j^-, \quad
x_i^- + \epsilon \le x_j^+,\quad
x_j^- + \epsilon \le x_i^+, & \text{if } (i,j) \in \mathcal{E}_v,
\end{cases}
\label{eq:lg:group}
\end{equation}
where $\epsilon > 0$ enforces a minimum overlap in the orthogonal direction.

\subsubsection{Boundary Constraints}
To ensure designated blocks lie on chip boundaries, we introduce linear equality constraints on block edges:
\begin{equation}
\begin{aligned}
    &x_i^- = x_{s}^h,  \quad v_i \in \mathbf{B}_{\mathrm{Left}}, 
    &x_i^+ = x_{t}^h,  \quad v_i \in \mathbf{B}_{\mathrm{Right}},  \\
    &y_i^- = y_{s}^v,  \quad v_i \in \mathbf{B}_{\mathrm{Bottom}}, 
    &y_i^+ = y_{t}^v,  \quad v_i \in \mathbf{B}_{\mathrm{Top}}.
\end{aligned}
    \label{eq:lg:boundary}
\end{equation}
Here, $x_{s}^h, x_{t}^h$ and $y_{s}^v, y_{t}^v$ represent the source and sink of HCG and VCG, respectively. Figure~\ref{fig:cg} illustrates the final constraint graph with details of constraints for each edge type.

\subsubsection{Handling Remaining Constraints}
The hard blocks $v_i \in \mathbf{V}_h$ have fixed dimensions: $w_i = \bar{w}_i, h_i = \bar{h}_i$. The predefined blocks $v_i \in \mathbf{V}_p$ have fixed geometric shapes: $(x_i, y_i, w_i, h_i) = (\bar{x}_i, \bar{y}_i, \bar{w}_i, \bar{h}_i)$. These quantities are treated as constants rather than optimization variables, thereby reducing the effective dimension and accelerating convergence.
Furthermore, all blocks within a multi-instantiation group $\mathbf{M}_k \in \mathbf{M}$ share the same variable ($w_k, h_k$) to satisfy the constraints, further reducing the scale of the variables.

\subsubsection{SOCP Formulation}
The complete SOCP-based legalization problem is formulated as:
\begin{equation}
    \begin{aligned}
        \min \quad & W + H                                                                                 \\
        \text{s.t.} \quad
                   & x_t^h - x_s^h \le W, \quad y_t^v - y_s^v \le H,                           \\
                   & x_s^h\ge 0, \quad y_s^v \ge 0, \quad x_t^h \le W_f, \quad y_t^v \le H_f, \\
                   & \text{Constraints \eqref{eq:non_overlap} -- \eqref{eq:lg:boundary}}.
    \end{aligned}
    \label{eq:full_socp}
\end{equation}

By minimizing the $W+H$, we obtain a convex proxy for area minimization. Problem \eqref{eq:full_socp} is a convex SOCP that can be solved to global optimal solution $\boldsymbol{v}^{lg}$ in polynomial time while satisfying all constraints, yielding a fully legal floorplan that strictly satisfies all constraints with improved area.

\section{Experimental}\label{sec:exp}
In this section, we present a comprehensive evaluation of the proposed floorplanning algorithm. Our implementation is developed in C++ and uses several libraries, including PyTorch~\cite{paszke2019pytorch} for optimization and MOSEK~\cite{mosek} for solving SDP and SOCP problem. All experiments are executed on a machine with an AMD EPYC 7763 CPU @ 2.45 GHz and 128 GB RAM.

\subsection{Experimental Setup}
We conducted our evaluation on the open-source FloorSet-lite benchmark suite~\cite{floorset}, which provides modern SoC floorplanning instances with design constraints, along with ground-truth references for comparison. By leveraging a reverse-engineering methodology that ensures near-optimality by construction, these benchmarks establish a definitive ground-truth that enables precise quantification of the optimality gap for any proposed solver~\cite{2025FloorSetChallenge}.
Furthermore, we fixed the positions of the IOs as specified in the benchmarks and treated the bounding box of IOs as fixed-outline, while setting the aspect ratio of the blocks to a range of $[1/3, 3]$.
The benchmark statistics are shown in Table ~\ref{tab:benchmark-info}.

\begin{table}[htbp]
  \caption{Statistics of the FloorSet-lite benchmarks \cite{floorset}}
  \label{tab:benchmark-info}
  \centering
  \resizebox{\hsize}{!}{
  \begin{tabular}{ccccccccc}
    \toprule
    Benchmark & $|\mathbf{V}_s|$ & $|\mathbf{V}_h|$ & $|\mathbf{V}_p|$ & \#IOs & $|\mathbf{E}|$ & MIB & Group & Boundary \\
    \midrule
    intel\_lite21  & 18  & 2  & 1 & 68  & 129   & \{7\} & \{3, 3, 3\}    & \{3, 3, 3, 6\}   \\
    intel\_lite30  & 26  & 2  & 2 & 93  & 239   & \{3\} & \{3, 4, 3, 3\} & \{5, 5, 6, 5\}   \\
    intel\_lite40  & 34  & 5  & 1 & 107 & 261   & \{4\} & \{5, 3, 4\}    & \{7, 5, 5, 7\}   \\
    intel\_lite50  & 45  & 3  & 2 & 124 & 286   & \{5\} & \{4, 4, 5, 4\} & \{6, 5, 7, 7\}   \\
    intel\_lite60  & 54  & 5  & 1 & 138 & 946   & \{4\} & \{5, 5, 5\}    & \{8, 6, 7, 6\}   \\
    intel\_lite70  & 59  & 7  & 4 & 197 & 836   & \{4\} & \{6, 5, 6, 7\} & \{8, 8, 6, 7\}   \\
    intel\_lite80  & 70  & 5  & 5 & 163 & 3370  & \{7\} & \{6, 7, 7\}    & \{9, 7, 6, 8\}   \\
    intel\_lite90  & 73  & 11 & 6 & 237 & 2127  & \{3\} & \{7, 6, 7, 7\} & \{10, 8, 6, 7\}  \\
    intel\_lite100 & 86  & 7  & 7 & 377 & 775   & \{5\} & \{9, 7, 8\}    & \{11, 11, 6, 6\} \\
    intel\_lite110 & 96  & 5  & 9 & 286 & 4251  & \{3\} & \{9, 8, 8\}    & \{10, 8, 7, 8\}  \\
    intel\_lite120 & 104 & 10 & 6 & 330 & 10580 & \{7\} & \{9, 10, 9\}   & \{15, 11, 7, 7\} \\
    \bottomrule
  \end{tabular}
  }
\end{table}

We compare our method against the state-of-the-art constraint-aware simulated annealing floorplanner PARSAC, along with two ablated variants of our method: \textbf{Ours w/o Init \& Incre} (without SDP initialization and without incremental strategy) and \textbf{Ours w/o Incre} (with SDP initialization but without incremental strategy). For all methods, we conduct 100 parallel trials and report the best result.
We evaluate the methods in terms of HPWL, area, runtime, and constraint violations.
\begin{table*}[htbp]
\centering
\caption{Experimental results on the FloorSet benchmarks}
\label{tab:updated_comparison_final}
  \renewcommand{\arraystretch}{1.012}
 \resizebox{\textwidth}{!}{
        \begin{tabular}{c cc c cccc c ccc c ccc c ccc}
            \toprule
            \multirow{2}{*}[-0.3em]{\textbf{Benchmark}}
                           & \multicolumn{2}{c}{Ground-Truth$^\dagger$~\cite{floorset_testsuite_lite}}
                           &
                           & \multicolumn{4}{c}{PARSAC~\cite{2024PARSAC}}
                           &
                           & \multicolumn{3}{c}{Ours w/o Init \& Incre}
                           &
                           & \multicolumn{3}{c}{Ours w/o Incre}
                           &
                           & \multicolumn{3}{c}{Ours}                                                                                                                                                                       \\
            \cmidrule{2-3} \cmidrule{5-8} \cmidrule{10-12} \cmidrule{14-16} \cmidrule{18-20}
                           & HPWL                                         & Area
                           &
                           & HPWL                                         & Area  & \#Viol$^\ddagger$ & RT (s)
                           &
                           & HPWL                                         & Area  & RT (s)
                           &
                           & HPWL                                         & Area  & RT (s)
                           &
                           & HPWL                                         & Area  & RT (s)                                                                                                                                  \\
            \midrule
            intel\_lite21  & 4.22                                         & 6955  &         & 5.28    & 7020  & 0 & 20.62  &  & 4.24    & 7241.89  & 6.62  &  & 4.26    & 7206.39  & 7.03  &  & 4.43    & 6988.60 & 11.66 \\
            intel\_lite30  & 17.86                                        & 9880  &         & 25.29   & 11988 & 0 & 32.19  &  & 18.86   & 10102.40 & 8.47  &  & 18.88   & 10101.13 & 10.47 &  & 18.17   & 9961.12 & 17.47 \\
            intel\_lite40  & 17.97                                        & 12650 &         & 30.57   & 13552 & 0 & 40.96  &  & 18.06   & 12883.33 & 9.06  &  & 18.23   & 12990.46 & 8.15  &  & 17.37   & 12739.58 & 21.63 \\
            intel\_lite50  & 18.95                                        & 19710 &         & 33.87   & 22475 & 0 & 51.66  &  & 19.42   & 20567.24 & 10.65 &  & 19.09   & 20352.49 & 10.37 &  & 18.48   & 19945.92 & 17.02 \\
            intel\_lite60  & 94.90                                        & 17584 &         & 177.21  & 20406 & 0 & 78.62  &  & 101.28  & 17888.93 & 11.50 &  & 97.89   & 17634.66 & 15.27 &  & 95.25   & 17435.39 & 23.44 \\
            intel\_lite70  & 71.98                                        & 22218 &         & 136.83  & 25116 & 1 & 79.10  &  & 78.24   & 23525.42 & 9.27  &  & 74.85   & 22818.31 & 16.29 &  & 74.18   & 22617.82 & 22.80 \\
            intel\_lite80  & 407.96                                       & 24957 &         & 832.21  & 29760 & 6 & 148.41 &  & 434.36  & 25468.18 & 14.54 &  & 436.98  & 25409.05 & 24.56 &  & 425.77  & 24905.64 & 35.80 \\
            intel\_lite90  & 246.75                                       & 28684 &         & 574.98  & 35893 & 4 & 141.60 &  & 266.72  & 30044.63 & 15.80 &  & 263.63  & 29769.26 & 24.61 &  & 255.19  & 28792.09 & 32.06 \\
            intel\_lite100 & 104.69                                       & 29394 &         & 242.60  & 36890 & 3 & 172.37 &  & 118.71  & 33306.45 & 18.75 &  & 107.86  & 32520.66 & 31.68 &  & 104.89  & 31674.44 & 38.89 \\
            intel\_lite110 & 295.21                                       & 30316 &         & 715.33  & 37466 & 6 & 212.65 &  & 314.49  & 30862.64 & 19.74 &  & 314.27  & 30696.94 & 30.37 &  & 306.42  & 30209.92 & 36.85 \\
            intel\_lite120 & 1357.00                                      & 34715 &         & 3065.15 & 43086 & 6 & 349.43 &  & 1447.08 & 35285.16 & 38.55 &  & 1451.93 & 34965.64 & 55.23 &  & 1418.14 & 34628.26 & 67.60 \\
            \midrule
            \textbf{Ratio}
                           & 0.98                                         & 0.99
                           &
                           & 1.90                                         & 1.16  & -       & 3.57
                           &
                           & 1.04                                         & 1.03  & 0.50
                           &
                           & 1.02                                         & 1.02  & 0.68
                           &
                           & 1.00                                         & 1.00  & 1.00                                                                                                                                   \\
            \bottomrule
        \end{tabular}
    }
\begin{flushleft}
\footnotesize 
    $^\dagger$ Ground-truth references are derived via reverse engineering and ensures near-optimality by design, thereby enabling quantification of the optimality gap for any method.\\
    $^\ddagger$ \#Viol denotes the number of constraint violations. The violation column for our method is omitted as it robustly achieves zero violations across all instances.
    
\end{flushleft}
\end{table*}

\subsection{Comparison with SOTA and Ground-Truth}
Table~\ref{tab:updated_comparison_final} summarizes the comparison results between our method, PARSAC, and the optimal references across the 11 FloorSet benchmarks. Notably, our method achieves zero constraint violations across all benchmarks, so the violation column is omitted for our method.

Compared with PARSAC, our method consistently achieves better solution quality on FloorSet, with an average reduction of 90\% in HPWL and 16\% in area. The quality gap further widens as the benchmark scale increases, indicating that our floorplanning algorithm remains effective in more strongly coupled constrained instances. 
Compared with the ground-truth references, the theoretical optima account for 98\% and 99\% of our achieved HPWL and area, respectively, which shows near-optimal floorplan quality while maintaining robustness across all tested cases.

Furthermore, our method also provides clear advantages in feasibility and efficiency. It satisfies all constraints on all 11 benchmarks and is the fastest method overall, with about 72\% average runtime reduction compared with PARSAC. In contrast, PARSAC starts to show nonzero violations once the benchmark scale exceeds intel\_lite60, with up to 6 violations on larger cases, and its runtime degrades more rapidly as problem size grows. This behavior is consistent with the algorithmic difference between the two paradigms: PARSAC relies on stochastic perturbation and cooling schedules, whose search efficiency deteriorates in large highly coupled constrained spaces, while our method uses smooth constrained optimization with SDP initialization and incremental augmented Lagrangian refinement for stable convergence and stronger quality runtime tradeoff.

\begin{figure}[htbp]
  \centering
  \includegraphics[width=\hsize]{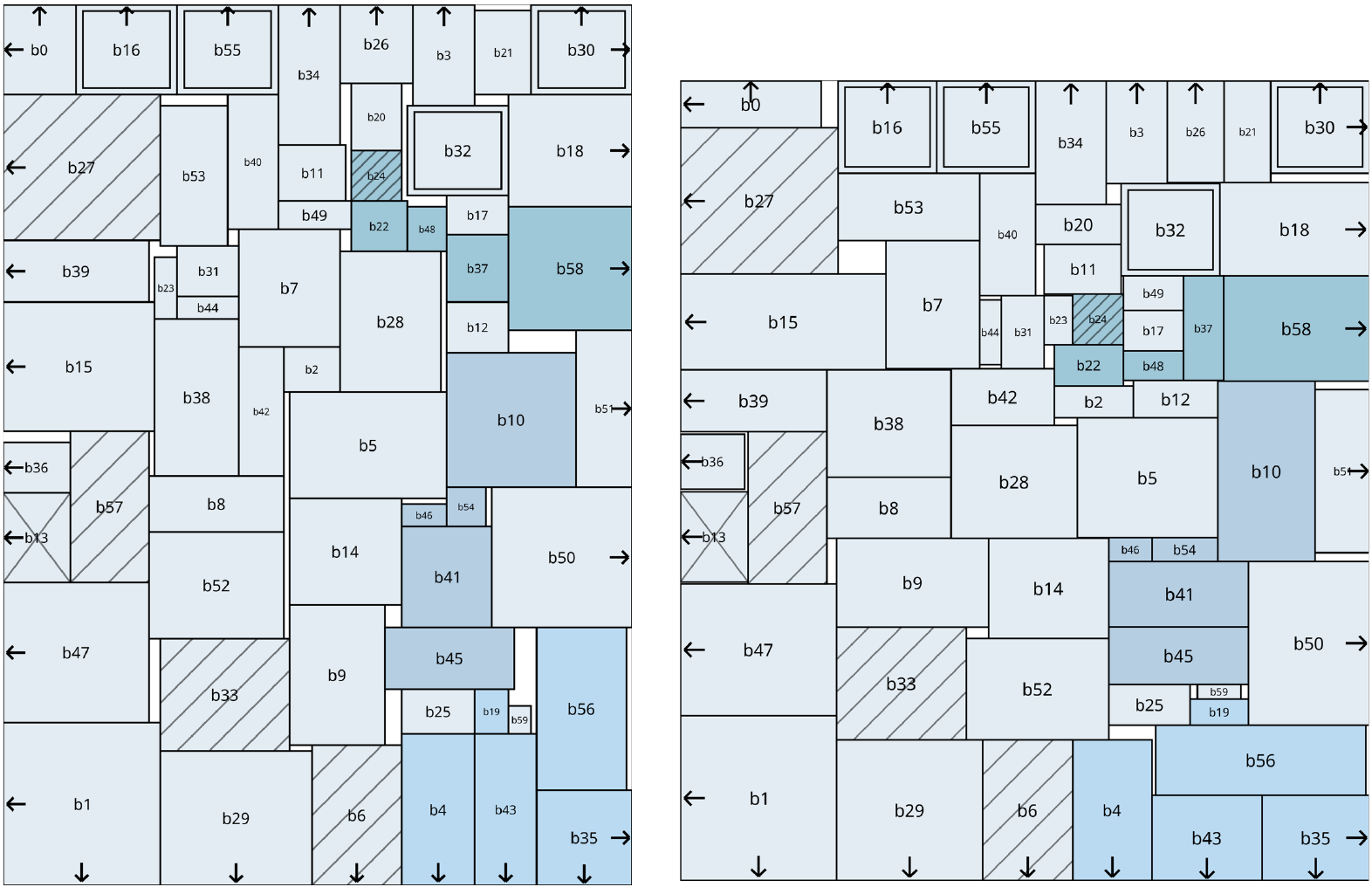}
  \begin{tikzpicture}[x=0.75pt,y=0.75pt,yscale=-1,xscale=1]
    \draw (0,0) node [anchor=north west][inner sep=0.75pt]  [font=\normalsize] [align=left] {};
    \draw (\hsize,5) node [anchor=north west][inner sep=0.75pt]  [font=\normalsize] [align=left] {};
    \draw (72,5) node [anchor=north west][inner sep=0.75pt]  [font=\normalsize] [align=left] {(a)};
    \draw (232,5) node [anchor=north west][inner sep=0.75pt]  [font=\normalsize] [align=left] {(b)};
  \end{tikzpicture}
  \caption{Visualization of experimental results on intel\_lite60 benchmark. (a) Ground-truth reference. (b) Ours.}
  \label{fig:ami49}
\end{figure}

Finally, to further demonstrate the near-optimality of our method, Figure~\ref{fig:ami49} presents a qualitative comparison between the ground-truth reference and our floorplanning result on intel\_lite60. Despite the highly non-convex nature of the problem, our method produces a strictly legal floorplan that closely matches the compactness and area efficiency of the reference solution, without relying on reverse-engineering techniques.

\subsection{Ablation Study}

The ablation results in Table~\ref{tab:updated_comparison_final} quantify the individual contributions of SDP-based initialization and incremental optimization. Compared with \textbf{Ours w/o Init \& Incre}, the full method achieves approximately 4.0\% reduction in HPWL and 3.0\% improvement in area. This result indicates that directly starting from a center-based initialization without subsequent refinement is computationally efficient, requiring only about 50\% of the full-flow runtime, but leads to a noticeable degradation in solution quality under coupled constraints.

Compared with \textbf{Ours w/o Incre}, the full method delivers an additional 2.0\% reduction in both HPWL and area. This confirms that the incremental optimization stage is essential for closing the remaining quality gap after SDP initialization. Although this stage increases runtime, the complete flow remains substantially faster than PARSAC, while consistently achieving better solution quality and robust constraint satisfaction.

\section{Conclusion}\label{sec:con}
This paper presents a novel analytical framework for modern constrained floorplanning, which unifies SDP-based initialization, incremental augmented Lagrangian optimization, and SOCP-based legalization into a coherent end-to-end solution. The proposed method effectively handles complex coupled design constraints while optimizing wirelength and area. Experimental results on the FloorSet benchmarks demonstrate that our approach outperforms the state-of-the-art by a significant margin in both quality and runtime, while strictly satisfying all constraints. This work opens up new avenues for applying advanced mathematical programming techniques to solve challenging problems in VLSI physical design. Future work includes exploring more efficient optimization algorithms, extending the framework to handle additional constraints, and applying it to larger-scale industrial designs.

\clearpage

\balance
\bibliographystyle{ACM-Reference-Format}
\bibliography{ref_copy}
\end{document}